# Adaptive Basis Sets for Practical Quantum Computing

*Hyuk-Yong Kwon,†¶ Gregory M. Curtin,†¶ Zachary Morrow,‡§ C. T. Kelley,‡ Elena Jakubikova†*

†Department of Chemistry, North Carolina State University, Raleigh, NC 27695, United States.

‡Department of Mathematics, North Carolina State University, Raleigh, NC 27695, United States.

§Scientific Machine Learning, Sandia National Laboratories, Albuquerque, NM 87123

Corresponding author e-mail: ctk@ncsu.edu, ejakubi@ncsu.edu

¶ These authors contributed equally

# ABSTRACT

Electronic structure calculations on small systems such as $H_2$, $H_2O$, LiH, and $BeH_2$ with chemical accuracy are still a challenge for the current generation of the noisy intermediate-scale quantum (NISQ) devices. One of the reasons is that due to the device limitations, only minimal basis sets are commonly applied in quantum chemical calculations, which allow one to keep the number of qubits employed in the calculations at minimum. However, the use of minimal basis sets leads to very large errors in the computed molecular energies as well as potential energy surface shapes. One way to increase the accuracy of electronic structure calculations is through the development of small basis sets better suited for quantum computing. In this work, we show that the use of adaptive basis sets, in which exponents and contraction coefficients depend on molecular structure, provide an easy way to dramatically improve the accuracy of quantum chemical calculations without the need to increase the basis set size and thus the number of qubits utilized in quantum circuits. As a proof of principle, we optimize an adaptive minimal basis set for quantum computing calculations on an $H_2$ molecule, in which exponents and contraction coefficients depend on the H-H distance, and apply it to the generation of $H_2$ potential energy surface on IBM-Q quantum devices. The adaptive minimal basis set reaches the accuracy of the double-zeta basis sets, thus allowing one to perform double-zeta quality calculations on quantum devices without the need to utilize twice as many qubits in simulations. This approach can be extended to other molecular systems and larger basis sets in a straightforward manner.

**INTRODUCTION**

Quantum computing has seen a recent uptick in investment and interest, largely due to the promise that quantum computational tasks could be completed exponentially faster than their classical counterparts.[1] One of the most direct applications for quantum computing is in the area of *ab initio* quantum chemical calculations[2,3] Particularly, the full configuration interaction (full CI) method, which is a nondeterministic polynomial-time complete (NP-complete) problem, is a difficult problem for classical computers as the dimension of the CI space increases factorially with the size of the basis set. For example, a full CI calculation for an $N_2$ molecule with ANO [4s3p1d] basis set needs $10^{10}$ Slater determinants requiring 80 GB of disk memory.[4] With the development of quantum circuit algorithms, full CI may become solvable even for large molecules within a reasonable time and resources available.[5-7]

However, the modern noisy intermediate-scale quantum (NISQ) devices are currently far too limited to accurately simulate large chemical systems. Many current quantum devices fall under this NISQ label, meaning they are incapable of large-scale error correction. Full error correction will only be achieved with at least several thousand quantum bits (qubits).[8] Given the current number of qubits available in a NISQ computer, quantum chemistry algorithms have been limited to simulating molecules with only a handful of atoms. Accurate modeling of small systems such as $H_2$, $H_2O$, LiH, and $BeH_2$ is an important step towards the use quantum devices for electronic structure calculations of larger systems, and develops a better understanding of the NISQ systems capabilities.[9,10]

While the structure optimizations on quantum computers are still challenging, quantum devices can be relatively easily employed to perform a series of single point energy calculations that allow one to construct a potential energy surface (PES) for a molecule of interest.[11,12] The PES can then be utilized to obtain useful information about the chemical system of interest, such as the molecular structure at a stable conformation (i.e., structures at PES minima), or a dissociation energy. The PES can be an important tool for modeling reaction dynamics or light-induced dynamics of chemical systems.[13-18] It can also be employed to perform vibrational analysis, allowing one to construct molecular partition functions and thermodynamic corrections to enthalpy and entropy of the system of interest.[19,20, 21] Therefore, even with the current NISQ limitations, one can start collecting useful chemical information with relevance to the molecular structure and reactivity from electronic structure calculations on quantum devices.

Due to NISQ limitations, minimal basis sets, such as STO-3G, are most commonly employed for quantum chemical calculations as the minimal basis set requires the smallest number of qubits.[9, 10, 22] The STO-*n*G basis sets possess a single basis function for each core and valence shell of a particular atom, composed of *n* primitive Gaussian basis functions, that were contracted to reproduce the shape of a Slater-type orbital.[23, 24] In the case of the STO-3G basis set, the s-orbital is described by a linear combination of three primitive Gaussian functions (assuming the atom is centered at the origin of the Cartesian system):

$$\Psi_{STO-3G}(r) = d_1 \left(\frac{2\alpha_1}{\pi}\right)^{3/4} e^{-\alpha_1 r^2} + d_2 \left(\frac{2\alpha_2}{\pi}\right)^{3/4} e^{-\alpha_2 r^2} + d_3 \left(\frac{2\alpha_3}{\pi}\right)^{3/4} e^{-\alpha_3 r^2} \qquad (1)$$

where $d_i$ are the contraction coefficients, $\alpha_i$ represent spatial spread of the wavefunction, and $r$ is the distance from nucleus. The STO-3G basis set is available for atoms from hydrogen to xenon.[25]

Traditionally, most basis sets are atom-centered and do not change with the molecular conformation. However, molecular orbital shapes and energies depend on relative atom positions. This is not a problem for large basis sets, but minimal basis sets do not have enough flexibility to describe the shapes of the molecular orbitals with the same accuracy at each molecular conformation. Thus, while the minimal basis sets are computationally efficient thanks to their small size, their accuracy can vary depending on the atom positions. On the other hand, the accuracy of complete or near-complete basis sets does not depend on molecular conformation, but they are computationally more expensive due to their larger size.[26, 27] For example, the calculated CISD/STO-3G energy of $H_2$ at 0.7 Å is 22.8 kcal/mol higher in energy than the energy calculated using a more complete basis set at CISD/aug-cc-pVQZ level, while the energy difference is 46.0 kcal/mol at 2.0 Å,[28] suggesting that the errors due to the basis set size are not uniform with respect to the molecular conformation (see Figure 1). Typically, one would increase the accuracy of the calculation by increasing the basis set size. For example, Figure 1 shows the significant improvement in both the potential energy surface (PES) energy and shape going from a minimal STO-3G and STO-6G basis sets to double-zeta basis sets (3-21G and 6-31G), and double-zeta polarized basis set (6-31G**).

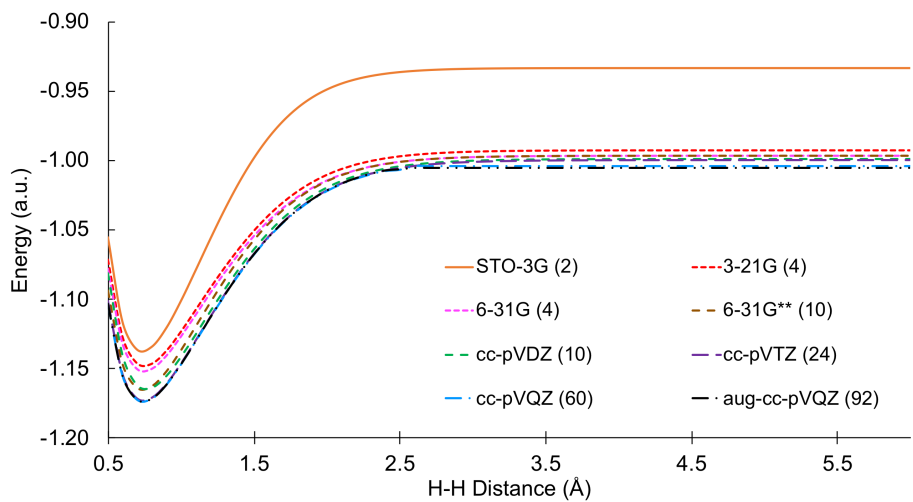

**Figure 1.** PESs of $H_2$ calculated at the CISD level of theory with different basis sets. Number of basis-functions utilized in each calculation is given in parenthesis after the basis set name.

Another way to increase the accuracy of small basis sets, without increasing the number of basis functions, is to change (or adapt) the basis set based on the molecular conformation, creating a so-called "adaptive" basis set. Adaptive basis sets employ basis functions with exponents and contraction coefficients that depend on the molecular structure. Decades ago, adaptive basis sets were employed to provide a better representation of the molecular orbitals, using the idea of extracting atomic orbitals from molecular orbitals with population analysis, which is useful for calculating electron correlation with quantitative *ab initio* formulation and the interpretation of reactions.[29-35] Adaptive basis sets were also utilized to reduce the computational cost of SCF algorithms.[36-41] Recently, Schütt et al. employed machine-learning to construct adaptive basis sets for large scale molecular dynamics simulations using polarized atomic orbitals and quasi-atomic minimal basis representation.[42-44] Due to the limitations on the number of available qubits in the current NISQ devices, as well as the increase in noise leading to substantial errors when using a larger number of qubits, adaptive basis sets provide an alternative way to increase the computational accuracy. Adaptive basis sets could improve the basis set quality without increasing the number of basis functions and thus the number of qubits utilized in the calculations, avoiding the additional noise and error that comes from using larger quantum circuits with bigger depth.

The aim of this work is to explore the use of adaptive minimal basis sets in quantum chemical calculations on NISQ devices, taking the $H_2$ molecule as our test system due to its small size and simplicity. We present an adaptive STO-3G basis set (labeled as A-STO-3G) that was constructed to produce the lowest energy at each H-H distance along the $H_2$ potential energy surface. Basis set contraction coefficients and exponents were optimized at the Hartree-Fock (HF), configuration interaction with single and double excitations (CISD), and density functional (B3LYP) levels of theory, resulting in three adaptive sets A-STO-3G_HF, A-STO-3G_CISD, and A-STO-3G_B3LYP.[45-47] All A-STO-3G basis sets are more accurate than STO-3G when compared to the benchmark aug-cc-pVQZ basis set. They show similar accuracy as unmodified STO-6G around the H$_2$ PES minimum, and a significant improvement over both STO-3G and STO-6G at long H-H distances. The PES of H$_2$ at the UCCSD/A-STO-3G_CI level of theory was also generated with several IBM-Q quantum devices to check its validity. The ability to produce accurate molecular PESs with quantum computers would help to extract useful chemical properties, such as dissociation energies and vibrational frequencies, and to enable studies of the chemical reactivity.

## METHODOLOGY

### Parameter optimization for adaptive basis sets

All calculations for basis set parameter optimization were carried out using the Gaussian 16 software package (Revision A.03).[48] Since HF and CI are variational methods, the basis set parameters can be optimized by minimizing the energy of the molecular system at each conformation. The pairs of $\alpha_i$ and $d_i$ parameters in Eq (1) are arbitrary, so we constrained $\alpha_1 > \alpha_2 > \alpha_3$ to obtain consistent results. Coefficient $d_3$ was set equal to one to avoid redundant results (sets of different coefficients and exponents that yield the same energy), brought upon by the normalization condition of the optimized basis function:

$$\int_0^\infty \psi(r)^*\psi(r)dr = \int_0^\infty N^2 \psi_{A-STO-3G}(r)^*\psi_{A-STO-3G}(r)dr = 1 \qquad (2)$$

where $N$ is the normalization constant and $\psi_{A-STO-3G}$ is the optimized basis function.

### Three-point optimizer

We have implemented the method of non-opportunistic coordinate search (see ref. 49 for a discussion of the classical coordinate search methods) in a script called three-point optimizer (TPO) for basis set parameter optimization with the Gaussian 16 software. A flowchart for the entire TPO process is found in Figure 2. TPO starts from the default STO-3G parameters at a given initial geometry. The optimizer then alters one coefficient by a user-defined constant, ± $c$, calculating single point energies at a set of parameters based on each alteration. If the center point is not the minimum of the three, the optimizer moves the center to the minimum among the three and repeats the step. The second or further iteration at this step requires only one more single-point energy calculation. If the center point between the three alterations is the minimum, then the optimizer moves on to the next coefficient until all five parameters (excluding $d_3$ set to one) yield optimization to the lowest energy, which is tested by another cycle through each parameter. Once all the parameters are optimized at the ± $c$ level, the process is repeated with a higher resolution ($c$/10) until it meets the criteria defined by the desired decimal place. At the next geometry, exponents and coefficients obtained in the previous step are employed as the starting point for optimization.

In addition to TPO, limited-memory Broyden-Fletcher-Goldfarb-Shanno algorithm with boundaries (L-BFGS-B),[50] and sequential least squares programming (SLSQP) optimizers[51] were tested. In the end, TPO optimized energies obtained were lower than energies from the two other optimizers but the computational cost of using TPO is much higher.

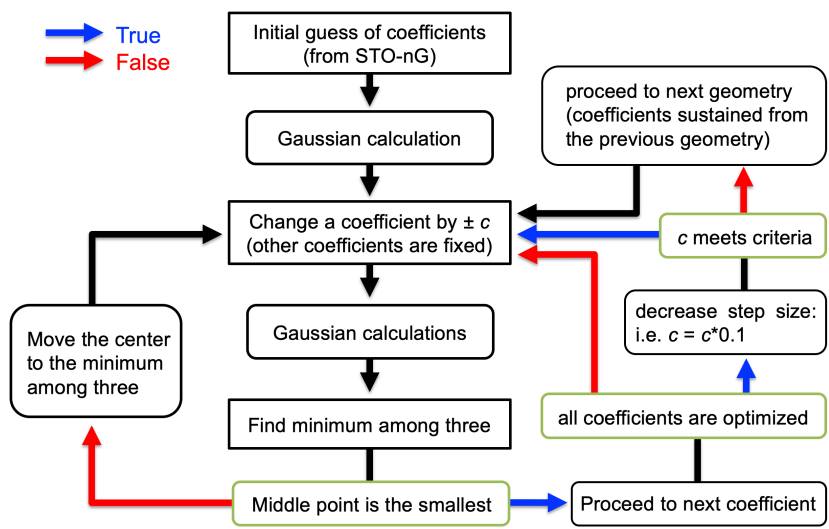

**Figure 2.** Flowchart of the TPO process.

**Sensitivity Analysis**

When constructing a basis set, it may not be known *a priori* whether all basis set parameters are equally important to determine the energy of the system. One or more parameters could be an order of magnitude more or less influential than the rest. Mathematically, the study of parameter influence is known as *sensitivity analysis* and is an active area of research.[52-58] Broadly, sensitivity analysis can be divided into two categories: local sensitivity analysis (LSA) and global sensitivity analysis (GSA).[59] LSA approximates sensitivities locally by evaluating the gradient of the quantity of interest (in this case, $E_0(r_{H-H}; \boldsymbol{p})$) at a particular parameter value.[59, 60] Frequently, LSA will utilize many evaluation points scattered across the domain to overcome the limitations of locality. Full GSA, on the other hand, constructs an $L^2$ expansion of the quantity of interest in terms of its conditional statistical moments.[61, 62] For basis-set construction, LSA was performed on each basis set coefficient parameter to identify which parameter is the most influential.

For ease of explanation and implementation, in this work we use LSA with $N = 200$ evaluation points of the parameters, $\boldsymbol{p} = (\alpha_1, d_1, \alpha_2, d_2, \alpha_3, d_3)$. The evaluation points are sampled from uniform distributions, which we chose because we must avoid negative $\alpha_i$ and because a uniform distribution is considered an unbiased prior distribution.[59] We also want to study parameter sensitivity as the H-H distance changes.

At a particular bond length $r_{H-H}$, we compute the sensitivity of parameter $i$ as

$$S_i(r_{H-H}) = \sqrt{\frac{1}{N} \sum_{j=1}^{N} \left( \frac{\partial E_0}{\partial p_i}(r_{H-H}; \boldsymbol{p}^j) \right)^2} \qquad (3)$$

where $\boldsymbol{p}^j$ is the *j*-th realization of the parameters. For the sampling intervals, we used

$$\alpha_1 \sim \mathcal{U}(1.5, 8.0), d_1 \sim \mathcal{U}(0.05, 0.15)$$
$$\alpha_2 \sim \mathcal{U}(0.3, 1.5), d_2 \sim \mathcal{U}(0.5, 1.0) \qquad (4)$$
$$\alpha_3 \sim \mathcal{U}(0.01, 0.3), d_3 = 1$$

which we determined by collecting the maximum and minimum optimized parameter values and then padding the relevant intervals even further. We need at least one $\alpha_i$ to be relatively small to yield a wide wavefunction, so we chose $\alpha_3$ without loss of generality. Furthermore, since the coefficients $d_k$ and $\alpha_k$ are normalized, we set $d_3 = 1$, again without loss of generality.

**H$_2$ potential energy surface construction**

Our adaptive basis set was developed over a range of H-H distances between 0.5 to 6.0 Å. The grid consists of equally spaced points at intervals of 0.1 Å within the specified range. The TPO is used at each grid point to find the parameters that minimize the energy at HF, B3LYP, and CISD levels of theory. The optimized basis set at each point can then be used in single point energy calculations at CISD level to construct the potential energy surface.

**Quantum device simulations**

The PESs of H$_2$ were generated by using qiskit 0.19.2 and qiskit-aqua 0.7.1 provided by IBM Quantum Lab paired with the PySCF driver.[63-66] For H$_2$ unrestricted singlet surfaces with STO-3G and A-STO-3G basis sets, the UCCSD variational form with the Bravyi-Kitaev mapping[67, 68] and COBYLA optimizer[69] were used for the variational quantum eigensolver (VQE) with 8000 shots.

**RESULTS**

In this section, we first discuss the results of the sensitivity analysis for the STO-3G parameters. Then we present the values of each optimized parameter across the entire H-H distance range. Next, we analyze the potential energy surfaces produced from these optimized parameters based on electronic structure calculations on classical computers. Finally, we report the results of the PES calculations using the optimized parameters on IBMQ quantum devices and simulators.

**Sensitivity Analysis Results**

We evaluated the sensitivity at the bond lengths 0.5, 0.6, …, 2.5 (in Å), equally spaced to every 0.1 Å, and the results are shown in Figure 3. The largest sensitivity corresponds to the exponent on the most diffuse Gaussian primitive, which is to be expected. However, the separation between the sensitivity of $\alpha_3$ and the other parameters is not several orders of magnitude, which is the typical heuristic of discarding parameters.[59] Additionally, the sensitivities have low variation as the H-H distance changes. Therefore, we optimize all six parameters, paying the most attention to $\alpha_3$ and the least attention to $\alpha_1$.

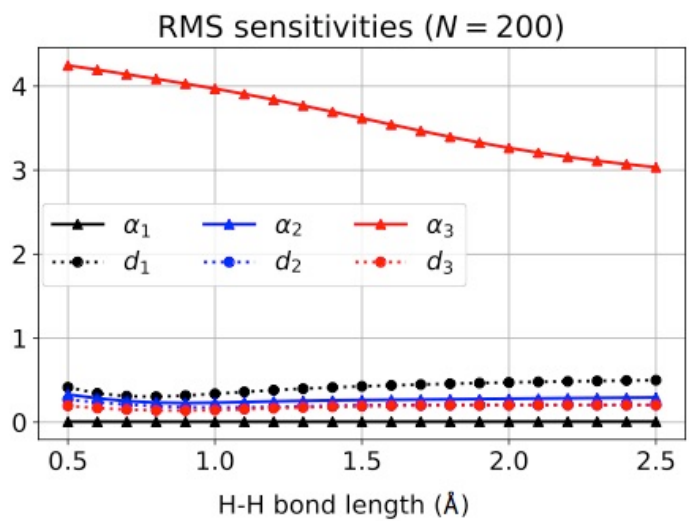

**Figure 3.** Sensitivity of energy to $\alpha_i$ and $d_i$ at different bond lengths.

**A-STO-3G Basis Sets: Parameters and Accuracy**

*Parameters for the A-STO-3G basis sets*

TPO optimization was done at the HF, B3LYP, and CISD levels of theory, resulting in the optimized A-STO-3G_HF, A-STO-3G_B3LYP, and A-STO-3G_CISD basis sets, respectively.

While the $H_2$ molecule is small and CISD calculations are relatively inexpensive, such calculations may not be practical for larger systems, where the use of HF or DFT methods may be more accessible and cost-effective.

The TPO optimized A-STO-3G parameters are plotted individually in Figure 4. As seen in Figure 4, optimized coefficients and exponents vary between the A-STO-3G optimized at the HF, B3LYP, and CISD levels of theory. For $d_1$ and $d_2$, the largest deviations occur at intermediate H-H distances. The value of $d_2$ at each level of theory returns to a near common value at longer bond lengths, while the CISD optimized $d_1$ remains lower than the other two theories. The optimized exponents $\alpha_1$, $\alpha_2$, and $\alpha_3$ show largest differences at longer H-H lengths. The PESs of $H_2$ calculated with A-STO-3G_HF, A-STO-3G_B3LYP, and A-STO-3G_CISD basis sets will therefore look similar at short and intermediate distances (< 2.5 Å), but are expected to show more significant variation at longer distances (> 2.0 Å). This observation is reflected in Figure 5, which displays the PESs calculated at the CISD level using both STO-$n$G and A-STO-3G basis sets. The HF-optimized PES was the highest energy of the optimized sets, most noticeable at longer distances, while the CISD-optimized PES energies were the lowest and closest to the more complete basis set at long H-H distances.

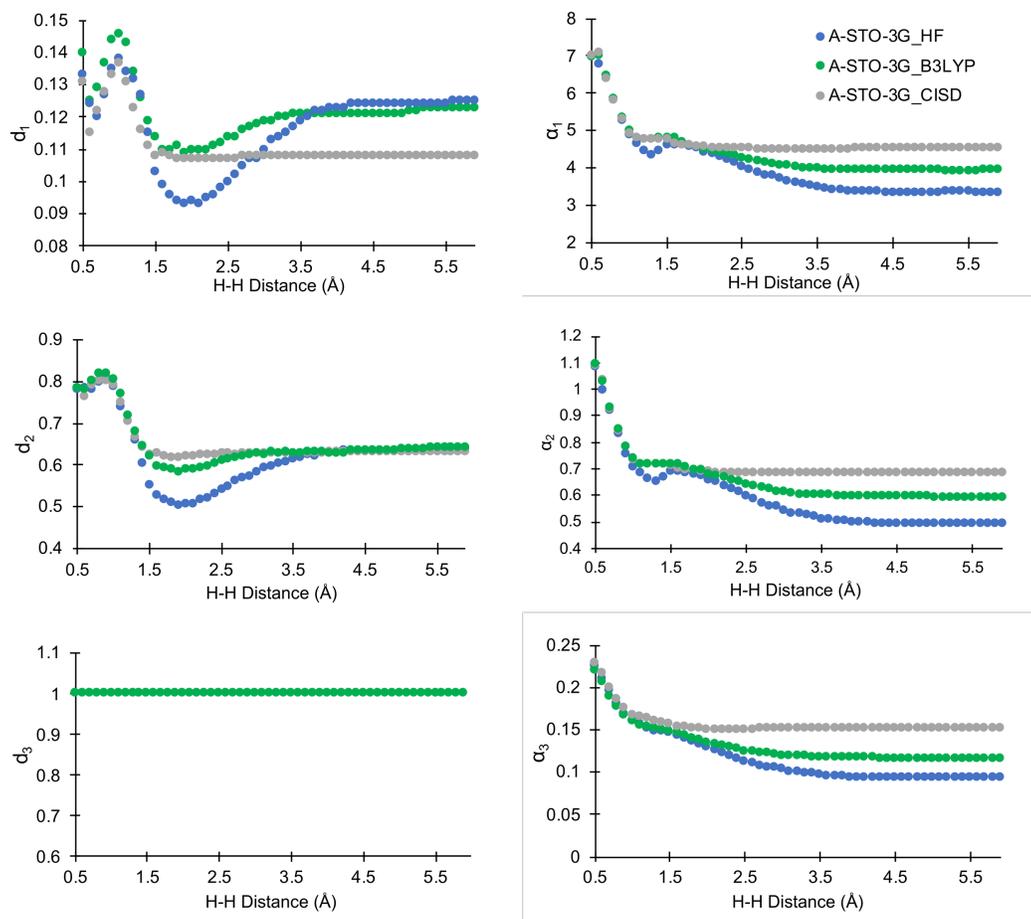

**Figure 4.** Optimized coefficients $d_1$, $d_2$, and $d_3$ (left) and exponents $\alpha_1$, $\alpha_2$, $\alpha_3$ (right) for all points in $H_2$ PES.

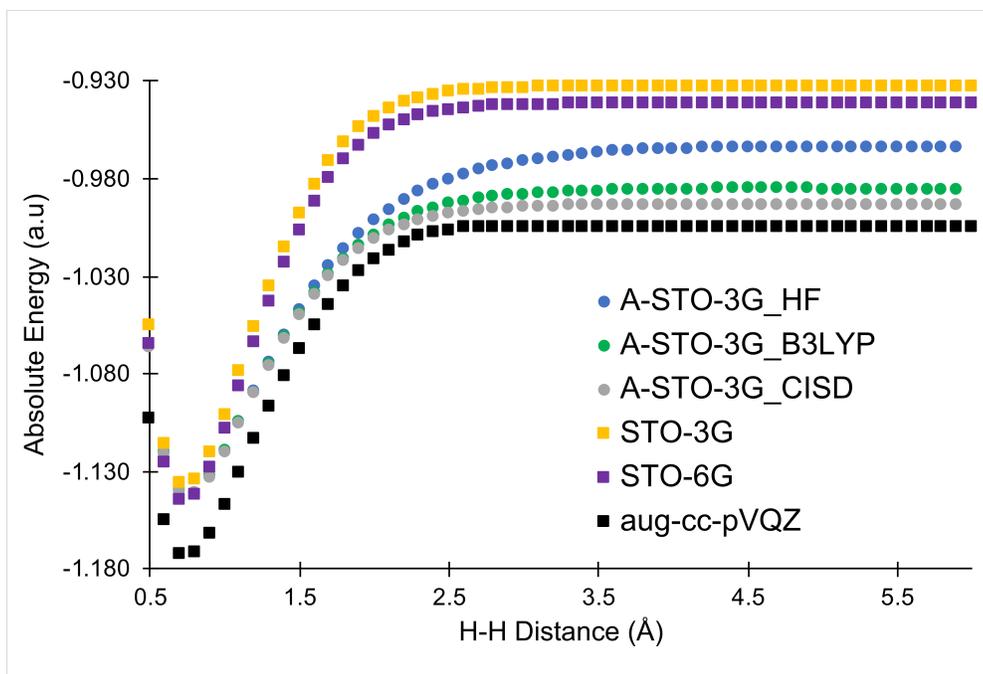

**Figure 5**. PES of H$_2$ at CISD level of theory with A-STO-3G basis sets obtained from the optimizations, along with the STO-nG and cc-p-VQZ basis sets for reference.

*Accuracy of A-STO-3G basis sets*

The PES of H$_2$ resulting from CISD/aug-cc-pVQZ calculations acts as the reference for determining if the use of the A-STO-3G basis sets has improved the accuracy. Neither the standard STO-3G, STO-6G, or the A-STO-3G variants match well with the reference around the minimum (H-H distance = 0.773 Å, see Figure 5). However, the energies obtained from the CISD/A-STO-3G calculations are significantly closer to the reference than the default STO-3G at long distances (see Figure 5, H-H distance > 0.9 Å). Importantly, the results obtained from the A-STO-3G_CISD basis set are very close to those obtained from the double-zeta quality basis sets (see Figure 6).

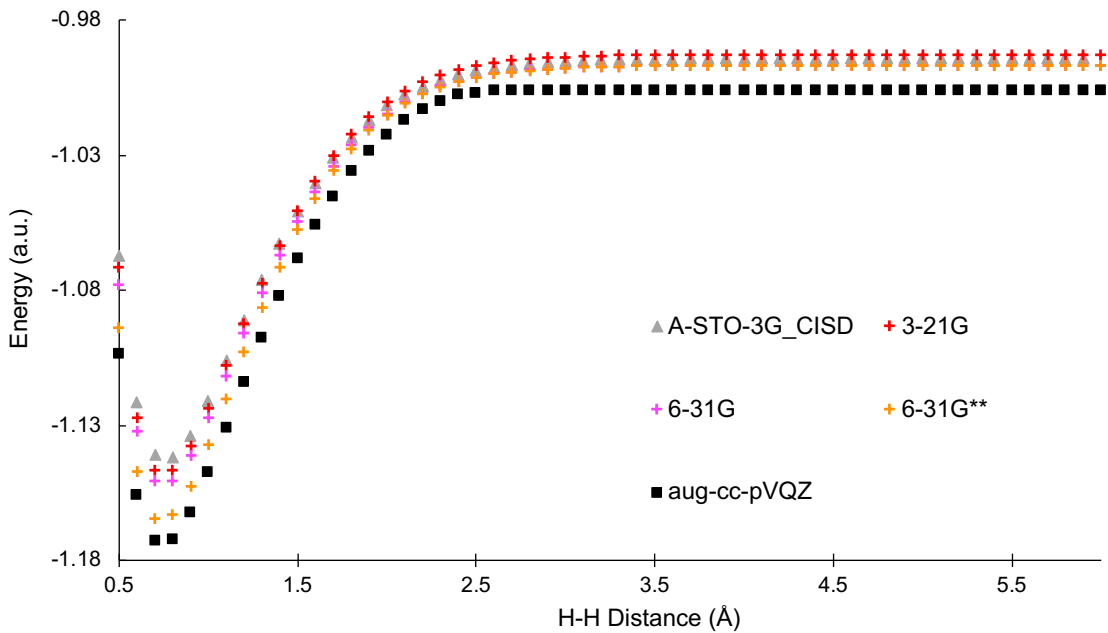

**Figure 6**. PESs of $H_2$ calculated at the CISD level of theory, comparing the A-STO-3G_CISD results with double-zeta basis sets and the aug-cc-pVQZ reference basis set.

Unmodified STO-6G is the most accurate at calculating the energies around the minima (H-H distance = 0.7 Å), followed by the A-STO-3G_CISD basis set that is ~3 kcal/mol higher in energy. For H-H distances longer than 1.0 Å, PES obtained using the A-STO-3G_CISD basis set demonstrates a considerable improvement in accuracy compared to the unmodified STO-3G and STO-6G; at H-H distance = 2.5 Å, the energy difference between the A-STO-3G_CISD basis set and aug-cc-PVQZ is only 7.27 kcal/mol, while the energy difference between the STO-6G and aug-cc-pVQZ basis sets is 39.85 kcal/mol. The PESs from HF- and B3LYP- optimized parameters also provide improved results, but are slightly less successful than the CISD-optimized STO-3G, leading to the errors at 2.5 Å of 25.46 kcal/mol and 12.29 kcal/mol, respectively. While it seems that it is preferable to employ CISD for the basis set optimization, parameters optimized at the B3LYP level provide similar accuracy to the CISD-optimized parameters, suggesting that basis set optimization utilizing DFT methods can be a good alternative for larger systems, where the use of CISD may become computationally expensive.

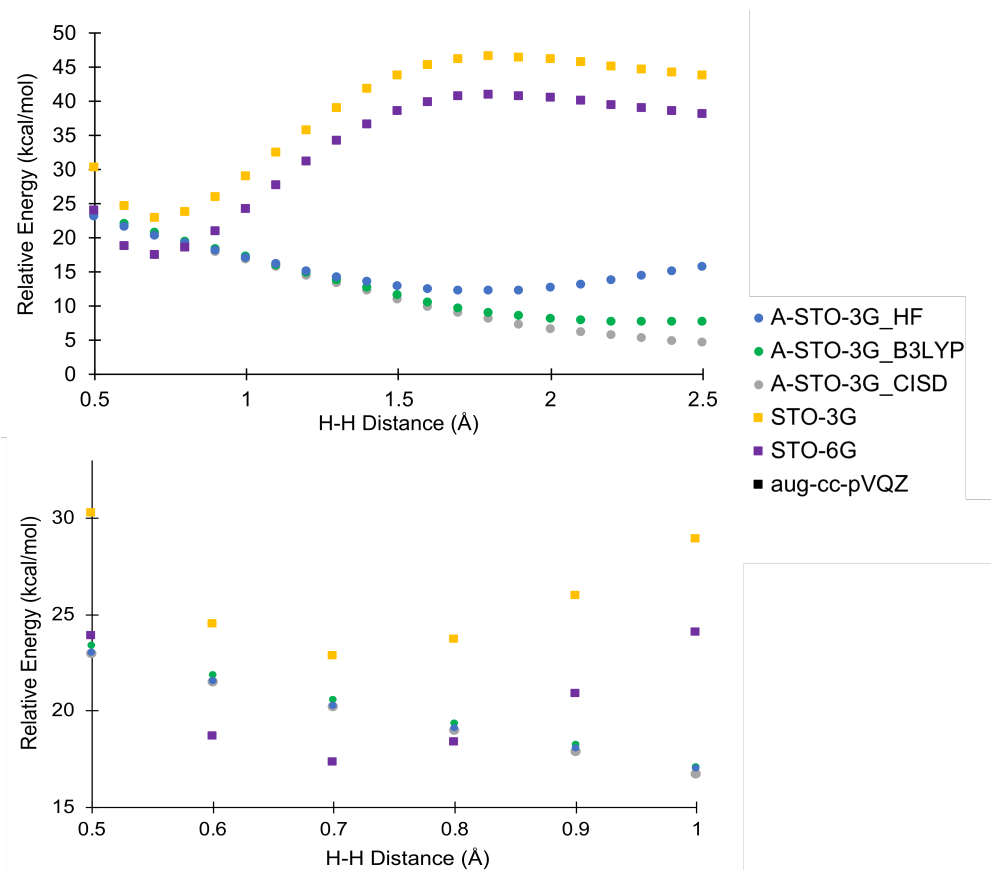

**Figure 7.** Top) Relative energies compared to the PES of CISD/aug-cc-pVQZ. Bottom) Close-up of 0.5 Å to 1.0 Å region.

Comparison of the calculated energies between the STO-$n$G, A-STO-3G and the reference aug-cc-pVQZ basis sets represents only one aspect of evaluating the basis set performance. The shape of the PES is another property that can be compared to the reference PES. Figure 7 (top) displays the PESs plotted on a relative energy scale, where the minimum of each PES is set equal to zero. The PESs calculated with the STO-3G and STO-6G basis sets are virtually identical and show a large deviation from the aug-cc-pVQZ surface, making the shape of the well narrower and overestimating the dissociation energy (see below). On the other hand, PESs obtained with the A-STO-3G basis sets result in the surfaces with wider wells, and lower dissociation energies than the reference.

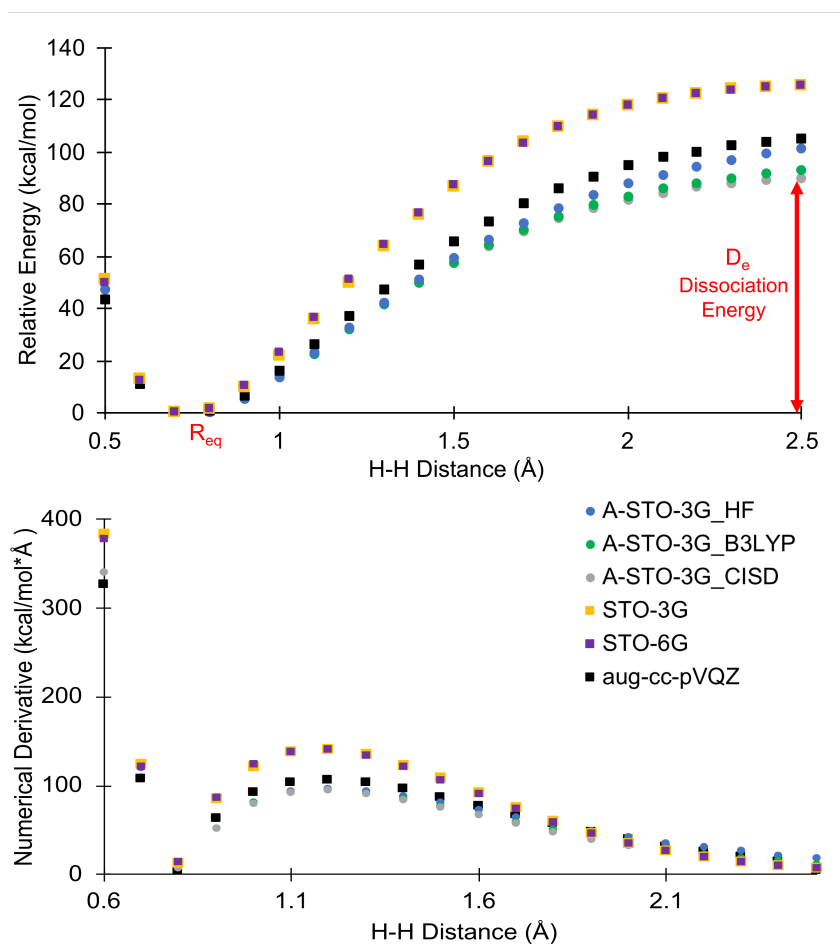

**Figure 8.** Top) Relative energies compared to minima of each PES. Bottom) Absolute numerical derivative of $H_2$ potential energy surfaces in kcal/mol·Å. $R_{eq}$ label stands for the equilibrium bond lengths, while $D_e$ denotes the dissociation energy.

The use of the A-STO-3G basis sets results in shapes closer to the reference, as illustrated by the numerical derivative of each PES calculated displayed in Figure 8 (bottom panel). The STO-*n*G basis sets perform worse compared to the numerical derivative of the reference until ~1.7 Å. In contrast, the numerical derivative of A-STO-3G_CISD PES is closer to that of aug-cc-pVQZ, meaning its overall shapes are more similar.

PESs can also be used to extract useful quantities such as the molecular structure at minimum ($R_{eq}$) or dissociation energy ($D_e$). We have obtained the minimum for each PES by first constructing the sixth-order polynomial fits and then calculating its first derivative and setting it equal to zero. Furthermore, the approximate dissociation energy for each PES curve was obtained as the change in energy from the minimum to the energy from a H-H distance of 6.0 Å.

Table 1 summarizes the obtained H-H distance at minimum as well as the H-H dissociation energy along with the experimental data.[70] Unoptimized basis sets provide more accurate minima, with STO-3G and STO-6G nearly identical to the experimental data. STO-$n$G basis sets were derived by least squares fitting of Slater-type orbitals which were designed to predict the correct atomic radius,[23, 24, 71] and in this case show greater accuracy than the CISD/aug-cc-pVQZ reference. The difference between HF, B3LYP, and the CISD optimized basis-set minima is roughly double that of aug-cc-pVQZ minimum. However, the $R_{eq}$ obtained from calculations with A-STO-3G basis sets are still reasonable and similar to those obtained from the double-zeta basis sets. Adaptive basis sets show improved PES accuracy over the unoptimized STO-3G at long H-H distance, which in turns improves the calculated dissociation energy.

The approximate dissociation energy is calculated as the change in energy from the point nearest the minima to the energy from a H-H distance of 6.0 Å. All dissociation energy values, including the experimental, are considered at 0 K. Our CISD/aug-cc-pVQZ reference shows high accuracy compared to the experimental result (1.54 kcal/mol). Optimized basis sets display between 56-76% increase in accuracy compared to the unoptimized CISD/STO-3G, and their values are similar to those obtained from the double-zeta basis sets.

**Table 1.** Calculated equilibrium distance and dissociation energy at CISD level of theory of optimized and unoptimized basis-sets. Experimental values taken from reference 70.[72]

|  | $R_{eq}$ (Å) | $\Delta R_{eq}$ (Å) | $D_e$ (kcal/mol) | $\Delta D_e$ (kcal/mol) |
|---|---|---|---|---|
| **Experimental** | 0.740 | 0.000 | 103.26 | 0.00 |
| **STO-3G** | 0.741 | 0.001 | 127.40 | 24.14 |
| **STO-6G** | 0.735 | -0.005 | 127.32 | 24.06 |
| **3-21G** | 0.795 | 0.055 | 96.58 | -6.67 |
| **6-31G** | 0.788 | 0.048 | 96.39 | -6.87 |
| **6-31G\*\*** | 0.763 | 0.023 | 105.10 | 1.84 |
| **aug-cc-pVQZ** | 0.773 | 0.033 | 104.80 | 1.54 |
| **A-STO-3G_HF** | 0.808 | 0.068 | 111.03 | 7.78 |
| **A-STO-3G_B3LYP** | 0.809 | 0.069 | 97.37 | -5.89 |
| **A-STO-3G_CISD** | 0.808 | 0.068 | 92.68 | -10.58 |

Overall, the use of A-STO-3G basis sets results in PESs that are slightly less accurate in approximating H-H equilibrium bond distance than the STO-3G basis set, but significantly more accurate for determination of the dissociation energies and the overall shape of the $H_2$ PES than the STO-*n*G basis sets. More importantly, the results obtained with the A-STO-3G basis sets reach the quality of the double-zeta basis sets that use twice as many basis functions. It is also worth noting that A-STO-3G basis set optimized at the CISD and B3LYP levels of theory provide very similar results, while the HF-optimized A-STO-3G is slightly less accurate.

**PESs from quantum devices**

The IBM Q Paris, Ourense, Rome, Bogota, and Santiago backends were used to collect data depending on their availability. The calculations were done at the UCCSD level of theory implemented in qiskit. Note that since hydrogen molecule contains only two electrons, UCCSD and CISD methods yield identical results. For the calculations on these quantum devices, parity mapping was used to reduce the noise by reducing the number of operated qubits to two. The SLSQP optimizer was used for VQE. The simultaneous perturbation stochastic approximation (SPSA) optimizer,[73] which is designed to work on noisy devices, showed more accurate results than SLSQP at longer distances. However, SPSA is about tenfold slower than SLSQP and becomes even slower for shorter distances requiring more optimization steps for energy to converge. Thus, we were not able to collect all data to construct the PESs with SPSA. With collected data from SLSQP, the trends in PESs between STO-3G and A-STO-3G basis sets are still evident.

We collected 395 data points for STO-3G basis set and 506 data points for A-STO-3G_CISD basis set from our runs on quantum devices. The PESs generated from the average and minima of the VQE results are presented in Figure 9. The results do not only deviate from the CISD surface, but also show a significant amount of noise even with 8000 shots per run. The major sources of error are likely from hardware such as qubit readout error, qubit instability, and CNOT (two-qubit gate) error. The magnitude of the error is not constant, but seem to increase at short and longer $H_2$ distances (see Figure 10). Based on the PESs fitted from the average values, it seems that the error could be interpolated with a continuous function (Figure 9), suggesting that it could potentially be mitigated by statistical analysis. Some of these errors can also be improved through recently introduced error-mitigation methods such as noise amplification[74] and classical postprocessing.[75] Overall, the data collected on the IBM-Q devices confirm the improvements in the $H_2$ PES at intermediate and long distances obtained through the use of the A-STO-3G basis sets that was seen on a classical computer, suggesting that development of the adaptive basis sets is a viable strategy to improve the accuracy of electronic structure calculations at quantum devices.

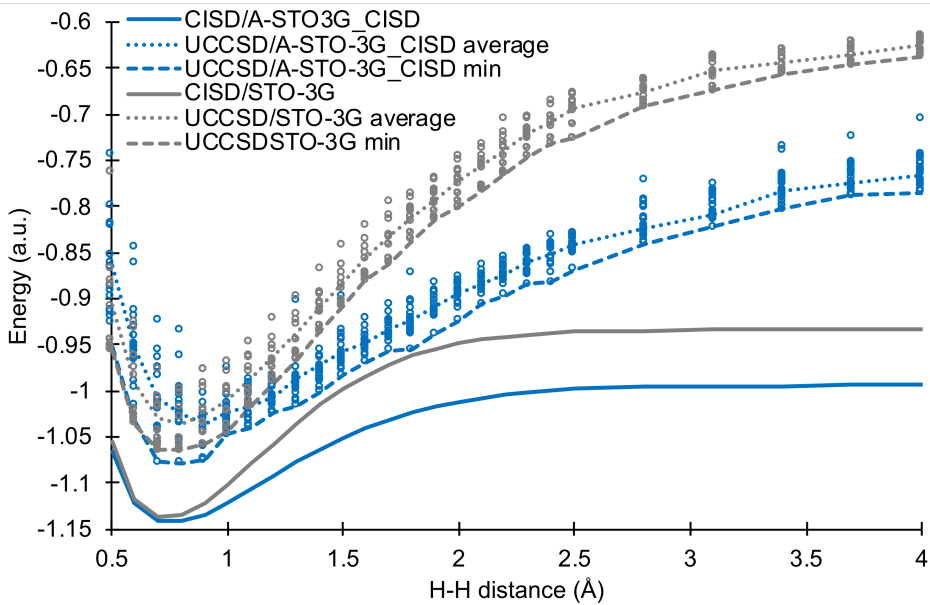

**Figure 9.** Results of $H_2$ simulations from quantum devices. Points represent individual $H_2$ energies obtained at the UCCSD level of theory with STO-3G basis set (grey) and A-STO-3G_CISD basis set (blue) from experimental runs on IBM quantum devices. Dotted lines represent PESs obtained from averaging energies obtained at each H-H distance, while dashed lines correspond to PESs obtained by considering the minimum energy at each H-H distance. The solid lines represent PESs of $H_2$ obtained at the CISD/STO-3G (blue) and CISD/A-STO-3G (grey) calculations on a classical computer. Note that CISD and UCCSD methods lead to identical results for $H_2$ calculations since the system contains only two electrons.

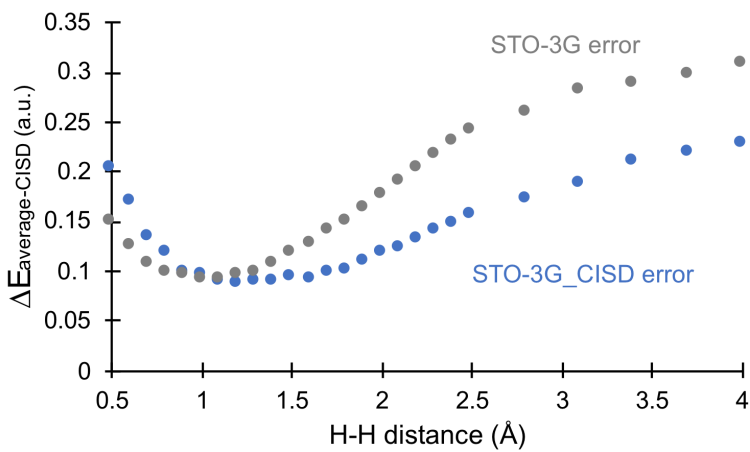

**Figure 10.** Energy differences between the averaged PESs obtained from UCCSD/A-STO-3G_CISD (blue) and UCSSD/STO-3G (grey) calculations on quantum devices and CISD/A-STO-3G and CISD/STO-3G calculations on a classical computer.

**Conclusions**

This work develops an adaptive basis set based on optimizing the STO-3G basis set for an $H_2$ molecule at multiple geometries. A-STO-3G basis sets are more accurate in calculating the $H_2$ PES than unmodified STO-3G. They are comparable to STO-6G around the PES minimum and lead to vast improvement at longer distances as the interaction between the two atoms diminishes. The A-STO-3G basis sets reach the accuracy of double-zeta basis sets (3-21G, 6-31G) that utilize twice as many basis functions. As a result, the PESs obtained with A-STO-3G basis sets possess a better shape and higher accuracy in reproducing the experimental dissociation energy. CISD/A-STO-3G_CISD calculations provide a PES closest in overall accuracy to the reference PES obtained at the CISD/aug-cc-VQZ level of theory, compared to the calculations with the STO-nG, A-STO-3G_HF or A-STO-3G_B3LYP basis sets.

The VQE calculations at the UCCSD/A-STO-3G_CISD level on quantum devices have a large error (> 0.1 hartree), but the average error may be corrected in post processing. Yet, the results obtained with A-STO-3G basis set still show a significant improvement compared to the STO-3G. Further statistical analysis on extracting useful chemical parameters such as vibrational frequency from noisy data will be performed in the future. Overall, adaptive basis sets provide a viable strategy to increasing the accuracy of electronic structure calculations on quantum devices without increasing the demands on the number of qubits or operations utilized in the simulations.

**Notes**

This paper describes work completed while the author ZM was at North Carolina State University. The authors declare no competing financial interest.

**Acknowledgement**

G.M.C, H.-Y.K, and EJ acknowledges support from the National Institute of Aerospace C15-2B00-NCSU-202023-NCSU and National Science Foundation CHE-1554855. C.T.K. acknowledges support from the National Science Foundation under Grants Nos. DMS-1745654 and DMS-1906446. Z.M. acknowledges support through an NSF Graduate Research Fellowship under DGE-1746939. The authors acknowledge use of the IBM Q for this work. The views expressed are those of the authors and do not reflect the official policy or position of IBM or the IBM Q team.